\documentstyle[epsf]{mn}
\def\sgast{${\rm Sgr\,A^*}\,$}

\begin{document}

\title[Circumnuclear Material]{The circumnuclear material in the 
  Galactic Centre:  A clue to the accretion process}

\author[R.H. Sanders] {R.H.~Sanders\\Kapteyn Astronomical Institute, 
P.O.~Box 800,  9700 AV Groningen, The Netherlands}

 \date{received: ; accepted: }


\maketitle

\begin{abstract}
On the basis of ``sticky particle'' calculations, it is argued that
the gas features observed within 10 pc of the Galactic Centre-- the
circumnuclear disk (CND) and the ionized gas filaments-- as well as
the newly formed stars in the inner one parsec can be understood
in terms of tidal capture and disruption of gas clouds on low angular
momentum orbits in a potential containing a point mass.  The 
calculations demonstrate that a dissipative component forms a 
``dispersion ring'', an asymmetric elliptical torus precessing counter
to the direction of rotation, and that this shape can be maintained
for many orbital periods.  For a range of plausible initial 
conditions, such a sturcture can explain the morphology
and kinematics of the CND and of the most conspicuous ionized filament.  
While forming the dispersion ring, a small cloud with low specific
angular momentum is drawn into a long filament which repeatedly
collides with itself at high velocity.  The compression in strong shocks
is likely to lead to star formation even in the near tidal field of the
point mass.  This process may have general relevance to accretion onto massive
black holes in normal and active galactic nuclei.

\end{abstract}
\section{Introduction}

Between two and five parsecs from the centre of the
Galaxy, there is a ring of neutral and molecular gas which
is usually referred to as the circumnuclear disk 
or CND (Genzel et al. 1994 and
references therein).  This feature is a clumpy, turbulent torus-like 
structure with a radial velocity field which is well-approximated by
rotation about the dynamical centre of the Galaxy.  Within the CND 
there is a central cavity where the overall gas density is significantly
lower, but where filaments of ionized gas are detected both in free-free
continuum radiation at radio wavelengths 
(Ekers et al. 1983, Lo \& Claussen 1983)
and in Ne$^+$ line emission at infrared wavelengths (Serabyn \& Lacy 1985).
The ionized gas
filaments, which may extend beyond the central cavity, have a very distinct
morphology-- sometimes called the ``mini-spiral''.  The total 
mass of the ionized gas is quite low-- less than 100 $M_\odot$.  
The radial velocity, as measured in the 
12.8 $\mu$m Ne$^+$ line (Serabyn et al. 1988) and in various 
hydrogen recombination lines (Schwarz et al. 1989, Roberts
\& Goss 1993, Herbst et al. 1993), 
indicate several distinct kinematic features having a 
systematic variation of radial velocity with position.  The most 
consipicuous of these features is the Northern Arm and its extention
to the west,  designated the ``extended Northern Arm'' by Serabyn 
et al. (1988);  it 
appears to wrap around the unresolved non-thermal radio continuum source,
\sgast which may be identified with a black hole having a mass possibly as 
large as $3 \times 10^6$ $M_\odot$.  

Also within the inner
one parsec there is cluster of young stars with a projected density which
increases to within a tenth of a parsec of \sgast (Genzel et.\ al 1996).  
This cluster constitutes a trivial fraction
of the mass of old bulge stars within this volume ($\approx 10^4\, M_\odot$
or about 1\% of the stellar mass, Morris \& Serabyn 1996), 
but contributes most of the
luminosity and all of the ionizing radiation in the central parsec.
In particular, this cluster of stars is the source of ionization for
the mini-spiral and the inner edge of the CND (Allen et al.\
1990, Eckart et al. 1993, Krabbe et al. 1995). 

There is now very strong dynamical evidence for a mass concentration in 
excess of
$2\times 10^6\, M_\odot$ centered at or near \sgast.  This is indicated
by the high velocities seen in the ionized gas filaments near \sgast  and
by the apparent increase in the measured radial velocites of the young
stars with decreasing distance from \sgast. The recent observation
of proper motion of the bright young stars in the central few arcseconds
(Eckart \& Genzel 1996)
make this conclusion almost inescapable;  one cannot appeal to an
anisotropic velocity distribution of these stars to account for
an increase in only the line-of-sight velocity dispersion.  But given
the existance of such a mass concentration, it is difficult to understand
the distribution and even the presence of the cluster of young stars 
(Allen \& Sanders 1986, Sanders 1992).
The estimated lifetime of the bright blue stars is on the order of a few
million years.  In the inner few tenths of a parsec this would correspond
to more than 100 orbit times, so one would expect that the cluster
should be thoroughly phase-mixed on this time scale.  Yet the appearance
is rather patchy and unmixed, with the centroid of the 20 or so most luminous
stars comprising the IRS 16 complex lying distinctly to the southeast of
\sgast by about one second of arc (Eckart et al. 1993).  
Moreover, the formation of stars within a few tenths of a parsec
of a massive black hole is quite problematic, because 
the density required for gravitational collapse in the
strong tidal field, the Roche limit, is 
$$\rho > 1.5\times 10^{-13} {{M_h}\over{10^6M_\odot}}
\Bigl({{0.1pc}\over r}\Bigr)^3\,\, {\rm g/cm^3} \eqno(1)$$
which is far greater than the observed or implied gas densities
in the central cavity.

It has been suggested that the ionized gas streamers may in fact be small
gas clouds on low angular momentum orbits which pass within 0.2 pc
of the central black hole (Ekers et al. 1983, Lo \& Claussen 1983,
Serabyn et al. 1988).  The idea is that such a cloud would be 
tidally stretched along its
orbit, and this has led to modeling the morphology and
kinematics of each of the various individual features
(see Fig.\ 2) by motion along Keplerian orbits appropriately
projected onto the plane of the sky (Serabyn et al. 1988, Herbst
et al. 1993, Roberts et al. 1996).  
It has been further suggested
that the young stars seen in this region may have resulted from
previous infall episodes;  the gas densities behind the strong shocks
expected in gas streams colliding with high velocity ($> 200$ km/s) 
could approach the Roche limit (Phinney 1988).

For understandable reasons, these original models for the gas filaments 
have been quite simple;  the
objective has been to provide a first order description of the observed
shape and motion of the features and to obtain limits on a central mass.  
Nonetheless, with such simple models, the kinematics and morphology of 
features such as the extended Northern Arm have been quite accurately 
reproduced (Herbst et al. 1993, Roberts et al. 1996).  On the basis
of this work it is
difficult to avoid the conclusion that the motion of the ionized gas
filaments is primarily orbital and that there is a large central mass
concentration.

Given the likely existance of a massive black hole at the Galactic Centre,
the structure and kinematics of the surrounding material-- the CND, 
the ionized filaments, and the cluster of young stars-- offers a unique 
close-up view of the accretion process which may be relevant
to galactic nuclei in general.   Here, using a sticky particle code, 
I consider this accretion process in the context of tidal 
disruption of clouds on low angular momentum orbits about
the dynamical centre.  The gravitational potential in which the gas clouds
move is that of a point mass and an extended spherically symmetric 
mass distribution similar to that of an isothermal sphere as suggested
by near-infrared observations of the stellar component in the central
region (Genzel et al. 1996).  Because of viscous dissipation
the tidal debris from such a cloud forms an elliptical
annulus of gas which precesses about the centre opposite to the sense of
particle motion.  

Such a structure can be described as a ``dispersion ring'',
to borrow an old term from galactic dynamics (Lindblad 1956, Oort 1965).
This is an ensemble of non-circular orbits in an arbitrary axisymmetric 
gravitational potential.  Each of these orbits is a closed
ellipse in some rotating frame;  in certain circumstances
it is possible to organize these orbits over some range
in energy or radius so that they precess with about the same angular velocity.
Then a non-axisymmetric structure, in this case an off-set elliptical
annulus, can be constructed in an axisymmetric potential-- a 
structure which may persist for a number of 
characteristic rotation periods (see Figs.\ 10a and 10c), 
as is the case for elliptical
accretion discs in a point-mass potential (Syer \& Clarke 1992).
Here the self-organization of the gas into such a dispersion ring
develops through the process of dissipation.

For a roughly spherical cloud with a radius of about 2 pc, initially located 
between 6 and 10 pc of the centre on an orbit which  carries it within
3 or 4 parsecs of the centre, the resulting dispersion ring is
broad and asymmetric with a central cavity of 1 to 2 pc-- similar in
structure and kinematics to the CND.  This similarity supports the
idea that the CND has resulted from the disruption of a cloud about
one million years ago.

At least some of the ionized
gas filaments in the central cavity may have a similar explanation, 
although in this case, a small cloud with a radius less than 0.5 pc
at an initial distance of 2 and 3 pc from the centre must pass within about
0.2 pc of the central point mass.  Then the resulting 
highly elliptical dispersion ring has a 
structure and kinematics very similar to that observed in the extended 
Northern Arm, and it persists for much longer than an 
orbital period ($\approx 5 \times 10^4$ years). 
On the first passage by the point mass, tides stretch the cloud into a
long filament.  On subsequent passages, while forming the dispersion ring,
the filament intersects itself at high velocity, and this can
lead to compression in strong shocks and probable star formation.  Thus
many of the young stars seen in this region may have resulted from this
or earlier such accretion episodes;  in any case, the gas now comprising 
the extended Northern Arm  may well be only a bare remnant of the 
original cloud.

All of this suggests that, in general, 
accretion onto massive black holes in the
nuclei of spiral galaxies may proceed by such tidal capture of low angular
momentum clouds.  If so, the accretion process is likely to be
highly episodic but also highly
inefficient with most gas disappearing in star formation rather
than being consumed by the black hole.  The essential ingredient is a 
clumpy, turbulent interstellar medium in the inner few hundred parsecs.
Such a medium is directly observed in the Galaxy and may be an 
attribute of the central regions of many spiral galaxies.

In all that follows the distance to the Galactic Centre is taken to
be 8.5 kpc.

\section{Calculations}

Earlier calculations of tidally stretched cloud models which take into
account gas dynamical effects include those of Quinn \& Sussman (1985)
and Bottema \& Sanders (1986).  Quinn \& Sussman
follow the motion of particles under the influence of gravity
and a drag force induced by an ambient medium.  In order to significantly
affect the motion of the material the ambient medium must have a density
approaching that of the filaments. 
Bottema \& Sanders carry out  a fully hydrodynamic simulation which 
suggests that a rather complex structure, not dissimilar to that
observed in the Galactic Centre, can form due to multiple passes of
material from a single cloud.  However, the technique applied here,
an Eulerian first order scheme with a square grid and transparent boundaries, 
is inappropriate to this problem (e.g., material leaving the grid can 
never return).  The ideal method is a sticky particle or SPH 
routine because motion of material can be followed arbitrarily close
to, or far from, the central point mass.  Moreover, the fact that orbital
motion is highly supersonic implies that the most important gas dynamical
effect will be dissipation is strong shocks rather than ordinary 
pressure gradient forces; it is unnecessary to apply full SPH to model the
essential effects (see Whitehurst 1988 for a discussion of this issue).

In the present calculation the gravitational potential
is that of point mass  embedded
in an extended mass distribution (the old stellar population) 
which makes a significant contribution to the
force at radial distances larger than about 1 pc.  The density 
distribution of the dominant stellar component is taken to be 
spherically symmetric and given by
$$\rho_*=\rho_o\Bigl(1+{r\over{r_c}}\Bigr)^{-2}. \eqno(2) $$
where the central density is $\rho_o = 2.67\times 10^{7}{\rm M_\odot
\, pc^{-3}}$ and the core radius is $r_c = 0.085$ pc.  When combined with a
$2.5\times 10^6$ $M_\odot$ point mass at r = 0, this yields an
mass distribution within the inner 5 pc which is identical, 
within the observational errors, to that derived by Genzel et al. (1996)  
on the basis of the Jeans equation using the measured projected stellar
density distribution and velocity dispersion.  
The total gravitational force in the inner few parsecs is then given
by $$ f  =  {{8.9\times 10^2}\over {r^2}}{\Bigl[{r\over {r_c}} 
- {\rm arctan}({r\over {r_c}})\Bigr]} $$ 
$$\hbox{\hskip 1cm} + {{4.3\times 10^3}\over{r^2}} \Bigl({{M_{bh}\over
{10^6 M_\odot}}\Bigr)}\,\,{\rm(km/s)^2pc^{-1}}. 
\eqno(3) $$
where $M_{bh}$ is the mass of the central object.  In all calculations
described below, $M_{bh} = 2.5 \times 10^6$ $M_\odot$
which is the value indicated by the stellar kinematics (Genzel et al. 1996), 

The calculations follow the motion of 4000 particles in the orbital plane;
i.e., the calculation is two dimensional.  This can be justified 
by full three-dimensional calculations (E. Hartlief, private communication 
1996) which  
demonstrate that when dissipation is included a spherical cloud of 
particles collapses very rapidly 
into the orbital plane on the first passage by the point mass 
(half the particles with positive velocity perpendicular to the plane
collide with the half traveling in the opposite direction).
The equations of motion for the particles are integrated by means of
a fourth-order Runge-Kutte technique with the time step continually
adjusted to achieve a specified level of accuracy.

The technique for including dissipation is similar to earlier methods
(e.g. Schwarz 1981, Whitehurst 1988, Jenkins \& Binney 1994) 
and combines aspects of sticky
particle and SPH algorithms.  In two dimensions every particle is a circle
with a radius $\sigma$ chosen to be sufficiently large 
such that each particle overlaps about 10 of its neighbours.  At any time
step every particle adjusts its velocity slightly to reduce the radial 
velocity
difference with each neighbour provided that the radial velocity difference
is negative (i.e., the particles are approaching).  
For two particles $i$ and $j$, we take
${\bf V}_{ij}$ to be the component of relative velocity along the line joining
the two particles (${\bf V}_{ij}$ is a vector at the position of particle
$i$ pointing away from particle $j$).  
If $r_{ij}$ is the separation between particle $i$ and $j$
then in time step $\Delta t_k$ particle $i$ changes its velocity
by an amount given by the vector sum
$$\Delta {\bf v}_{ik} = \alpha_k\sum_{j}^{r_{ij}<\sigma}{{\bf V}_{ij}}
\eqno(4)$$
where $$\alpha_k = \Delta t_k/\Delta t_s\eqno(5)$$
The strength of the interaction is proportional to $\alpha_k$.
Here $\Delta t_s$ is a standard time interval (a dissipation time scale)
which is typically taken to be 0.04 of a characteristic orbit time 
in the inner few parsecs ($\Delta t_s \approx 4000$ years).  

This is an algorithm for
including a bulk viscosity in which every particle's velocity is 
adjusted proportionally
to the local velocity divergence, but only if that divergence is negative
(i.e., the flow is converging).  The interaction strength, $\alpha_k$
depends upon the duration of the time step.  If this were not the 
case, then,  when considerations of numerical 
accuracy force the time step to be very short (as when several particles
are near the point mass), the entire fluid would become unrealistically
more viscous.  

In the calculations shown the motion of two interacting particles is not
effected by the transverse component of their relative velocity; that
is to say, shear viscosity is explicitly excluded.  This is done because
the primary mechanism determining the evolution of the debris is
dissipation, and this is adequately modeled by bulk viscosity.
Shear viscosity, when included, has no significant effect on the results,
but it does tend to make a fluid which is already quite possibly too
sticky even more so.

It is evident that inelastic interaction described by eq.\ 4
conserves the linear momentum
of two interacting particles. The total angular momentum of the 
ensemble of particles is also conserved; although, 
the angular momentum may be redistributed
among particles while the total energy decreases.  Griding is used to speed
up the search for neighbours, but, because of the
circular neighborhood, the interaction is isotropic
and bears no imprint of the imposed Cartesian grid.  As is typical in
such calculations, there is some tendency for the particles to 
bunch together;  this tendency is greater for smaller values of
$\Delta t_s$.  Nonetheless, the gross features of the calculation turn out
to be rather insensitive to the exact values of $\sigma$ and 
$\Delta t_s$ provided
that each particle actually overlaps several other particles at any given time.

\section {Simulation of the CND}

The CND as 
observed in HCN, together with the ionized gas filaments as seen
in the 6 cm continuum, is shown in Fig.\ 1 which is reproduced from
the paper of G\"usten et al. (1987).
In a recent review, Morris \& Serabyn (1996) emphasize the asymmetry
of the CND and its probable transient nature.  From inspection of Fig.\ 1
it is evident that \sgast, presumed to be at the dynamical centre of the
Galaxy, is not at the centre
of the central cavity of the CND but is closer to its western edge.  
Moreover, the molecular gas is significantly more tenuous on the 
eastern side of the center.  The
ionized gas associated with the inner edge of the CND is also asymmetrically
distributed;  the free-free emission does not form a complete ring
but only appears as the
Western Arc of the mini-spiral.  The outer part of the CND is clearly 
asymmetric and more extended to the south than to the north.  
All of this suggests
that the CND may be formed by capture and disruption of a passing cloud,
and this is the possibility pursued here.

The simulation is that of a clumpy
cloud on a low angular momentum orbit in the gravitational field
described above (eq.\ 3).  The initial radius of
the cloud is approximately 2 pc and it is located at 6.5 pc
from the Galactic Centre.  The cloud consists of 40 sub-units (clumps)
each comprised of 100 particles, and the velocity dispersion between
the clumps is 40 km/s.  The clumps have a systematic velocity only in the
tangential direction which may be expressed in terms of the circular 
velocity $V_c$-- the velocity required for a circular orbit at this
radius--
$$V_t = fV_c.\eqno(6)$$
Here f=0.4 or $V_t\approx 40$ km/s.  
Therefore, the cloud will fall toward the dynamical 
centre.  Such an initial configuration might result from the collision
of two oppositely moving clouds, not an unlikely event in a highly
inhomogeneous and turbulent medium.  Given that the mass of the CND
is roughly $10^4$ $M_\odot$ (Mezger et al.\  1996), the mean
particle density of the initial cloud is approximately $10^4$ ${\rm cm^{-3}}$;
the cloud would be stable against gravitational collapse if
the velocity dispersion exceeds 5 km/s.

The 4000 sticky particles are characterized by a radius $\sigma = 0.1$ pc,
and an interaction time interval
of $\Delta t_s = 4000$ years (eqs.\ 4 \& 5).  This means that velocity
of two interacting particles can change by as much as their 
velocity of approach over 4000 years.  The duration of the time step 
typically varies
between 50 and 250 years, so the actual change over a time step in
the velocity of two interacting particles is only a few percent of
the approach velocity.

The results are shown in Fig.\ 2 which is a
time sequence showing the passage of this cloud in the orbital plane.
Each frame is a snapshot of the morphology of the cloud at the indicated
time, in units of a million years, since the beginning of the infall. 
The cloud, on its first passage by the point mass, is stretched by
the tidal force into a long filament.  Subsequently, the filament
wraps about the centre and collides with itself once per orbital period
in the first two or three periods
(note that the orbit time at 3.5 pc from the centre is about $2\times 10^{5}$
years).  After two or three characteristic orbit times 
($\approx 5 \times 10^5$ years), the cloud has
settled into an asymmetric circumnuclear ring with a central cavity.
If the force were strictly that of a point mass the tidal 
debris would form a Keplerian ellipse, fixed in the orbital plane,
which would persist for many orbital periods.  Indeed, in such a calculation
with a $1/r^2$ force law, the distribution of particle orbits
show no tendency for circularization over 10 characteristic orbital periods
which is consistent with the results of Syer \& Clarke (1992) on Keplerian
elliptical accretion discs.  However, in the case considered here,
because of the contribution of the Galactic potential the force is
not $1/r^2$.  This means that different parts of the elliptical accretion
disc precess at different rates, and the process of circularization is more
rapid than in the Keplerian case.  But during this process, the tidal
debris temporarily forms a broad elliptical
ring with a one-arm spiral 
which is precessing counter to the sense of particle motion-- 
a ``dispersion ring'' as defined in the Introduction.  The ring 
is fully developed 
after a million years, and gradually, over two or three million years,
the central cavity fills in and the ring contracts while becoming more
circular.

The tidal debris, as it appears in the seventh frame of Fig.\ 2 (t=0.85),
provides a fair description of the CND.  Rotating this structure 
clockwise by 130 degrees, tilting the positive y-axis forward
about the x axis by 60 degrees (the
inclination of the orbital plane) and rotating  
the negative x axis (the line-of-nodes) to a position angle 
of 35 degrees with respect to the vertical (north), the ring would
appear as shown in Fig.\ 3a which is on the same scale as Fig.\ 1.  
These are typical projection parameters for the CND where the northwestern
side is assumed to be the near side.
The arrows indicate the magnitude and direction
of the velocity projected onto the plane of the sky and the ``X'' marks 
the position of \sgast; the straight line is the line-of-nodes.  
A contour map of the gas surface density is shown in Fig.\ 3b.
It is evident that the size and the asymmetry 
of the CND are approximately reproduced.  The two dimensional
radial velocity field is shown in the form of a contour map in Fig\ 3c,
and the radial velocity at 
the inner edge of the ring as a function of position angle is compared 
to the observations of G\"usten et al. (1987) in Fig.\ 4.  

In both morphology and kinematics
the agreement between the observations and model is reasonable.
In the model, the densest
part of the ring with the sharpest inner boundary is the western
side, nearest to the point mass;  the eastern side 
of the ring is rather tenuous and ill-defined.
This is due to the fact that the elliptical orbits in the dispersion ring
crowd together at the peri-centre.
Looking back at Fig.\ 1 we see that the distribution of molecular gas has
just this asymmetry.  This is not just an aspect of the particular 
simulation;  in the dispersion
ring model of the CND, that part of the inner boundary which is closet
to the dynamical centre, will always have the highest gas density due to
the orbit crowding.  
Moreover, because of the higher gas density along the western rim of the
cavity, it is likely that the emission measure of the ionized gas
is greater here than elsewhere along the inner boundary; that
would be consistent with the fact that only this part of the cavity
boundary reveals itself as an ionized filament (the Western Arc).  

The model considered
here, with the adopted orientation, reproduces the observed outer asymmetry
of the CND, the extention to the south, because of the one-arm 
spiral feature.  In the projection
required for the inner edge asymmetry, this feature does extend to the
south provided that the near side of the CND is the northwestern side. 
However, an aspect of the observed morphology which is not
reproduced by this simple model is the extention to the northwest.  
This feature might
result from initial irregularities in the shape of the cloud or from
a separate infall event (e.g., associated with the Northern Arm).

It seems significant that the same model and orientation parameters which
reproduce the overall morphological aspects of the CND also give the 
closest match
to the observed kinematics.  In particular, the ridge of positive velocity
material to the west of \sgast (Fig.\ 3c) 
is a conspicuous aspect of the observations
(G\"usten et al. 1987) but would not be a property of an axisymmetric
disc.  The asymmetry in the observed kinematics of the inner edge
(Fig.\ 4) has been attributed to the warping of the plane of the
CND.  However, the model here reproduces these asymmetries via gas motion
on offset elliptical streamlines.  

All of these distinct 
aspects of morphology and kinematics--  the outer extention to the 
south, the highest density edge on the western side,
the kinematic asymmetry of the inner boundary, the appearance of
the positive velocity ridge to the west--
can, for the observed sense of rotation, only be 
reproduced for a specific orientation
of the model CND in space;  that is to say, the near side 
is the northwestern
side in the context of this scenario.  This is because the asymmetric
gas distribution and kinematics break the usual nearside-farside degeneracy
of axisymmetric structures.
This is a prediction which is independent of the details of
the orbital initial conditions; the consequence is that the proper motion
vectors of the gas must be in the sense shown in Fig.\ 3a.

Any calculation in which a
cloud with a size of two or three parsecs having an initial  galacto-centric
distance of six to eight parsecs and a tangential velocity about half
that required for a circular orbit will lead to roughly the same 
structure.  The essential-- and model-free-- results are that the tidal 
debris will form an elliptical and asymmetric dispersion ring which 
persists for at least a million years and that the central cavity
is also asymmetric with the side nearer the point mass having a
sharper inner boundary and higher gas density.  This structure,
with a very specific orientation in space, provides
a plausible representation of the CND.

\section{The ionized filaments}

The ionized filaments in the central cavity of the CND
are shown in greater detail in Fig.\ 5.
The most conspicuous and coherent of these features is the 
extended Northern Arm
which has a radial velocity that varies systematically with position 
over 30 seconds of arc as it sweeps around \sgast.  Due to the large gradient
in radial velocity near \sgast, it is certain that this feature actually
does pass within about 0.2 pc of \sgast-- a true distance 
which is comparable to the projected
distance (Herbst et al. 1993, Roberts et al. 1996).  
The structure of the Northern Arm is clearly delineated in the
12.4 $\mu$m continuum radiation from hot dust seen in Fig.\ 6
(Gezari, Dwek \& Varosi 1996).  The ``Eastern Arm'' appears less
conspicuous in this mid-infrared emission, and 
it is now quite clear that the 
``Western Arc'' is the ionized inner edge of the circumnuclear ring
(Lo \& Claussen 1983, Roberts \& Goss 1993).
Features such as the extended Northern Arm may result from the same process
as that giving rise to the CND.  If so, then because this filament 
passes so near \sgast, the 
specific angular momentum of the original cloud ($f$ in eq. 6)
must be smaller than that required for formation of the CND. 

Fig.\ 7 is again a time sequence of the evolution of tidal debris
resulting from an passage of a small cloud (a ``cloudlet'') 
on a very low angular
momentum orbit.  Initially the cloudlet is round with a radius of
0.4 pc, and the centre of the cloudlet is at a distance of 2.4 pc
from the Galactic Centre, moving tangentially (counterclockwise)
with f=0.2 in eq.\ 6 (about 20 km/s).  In this case the cloudlet has zero
internal velocity dispersion.  The interaction properties of the 
sticky particles are $\sigma=0.02$ pc and $\Delta t_s =
4000$ years.  The linear scale of the frames in Fig.\ 7
is a factor of three smaller than those of Fig.\ 2 and
the indicated times, again in units of 
$10^6$ years, are a factor of 10 
shorter than in the case of the CND simulation due to the close approach
of the cloud particles to the point mass.
Such a cloudlet could be a particularly low angular momentum clump
in the larger cloud forming the CND or it could result from a subsequent
accretion event.  The difference in time scales would seem to argue in favor
of the latter, but more will be said about this below.  If the mass of this
original cloudlet were comparable to that of the ionized gas in the
filaments ($\approx 100$ $M_\odot$), then the original density 
would be on the order of $10^4$ cm$^{-3}$ as in the cloud forming the CND.
Overall this cloudlet would have the properties of an observed clump in
the CND (Genzel et al. 1994).  Such clumps are subject to tidal disruption
at distances between two to three parsecs from the center (eq.\ 1).  It may be
that the clumps maintain their integrity by turbulent pressure (collisions)
and that occasionally such a collision will lead to the sort of low
angular momentum cloudlet required here (the clumps may also be denser
as suggested by the analysis of Jackson et al. 1993).

As in the previous simulation the cloud is tidally stretched into a long
filament which repeatedly plunges toward the point mass, colliding
with itself as in the fourth frame (t=0.075).  By $2\times 10^5$ years
a fairly coherent and relatively long-lived elliptical dispersion ring 
is formed which precesses counter to the direction of rotation 
with a angular velocity of $\Omega 
\approx 26\,\, {\rm km/s\,pc^{-1}}$ (i.e., the precession period is about
$2.5\times 10^5$ years).  The time
for this structure to settle into a more axisymmetric accretion disk
depends upon the viscous dissipation but may be as long as $10^6$ years.
This circularization time-scale, in terms of orbital periods, is longer
than that of the CND cloud because of the smaller initial size of the cloud;
there is less differential precession over the narrow dispersion ring.

On the way to forming the dispersion ring,
streams of gas collide repeatedly with high velocity, as at 
t = 0.10.  The structure at these epochs is not unlike 
that seen in the ionized gas streamers, and it is tempting to identify
this morphology with that of the three arm spiral.  Indeed with the 
appropriate projection the narrow filament can be matched with the
Northern Arm, Eastern Arm and Bar.  However, such a projection 
fails to reproduce the observed dependence of radial velocity
on position, particularly in the well-studied extended Northern Arm.
Both morphologically and kinematically, this feature
appears to be distinct from the Eastern Arm (see Fig.\ 6), and
the two should probably not be modeled as part of the same structure
(Roberts \& Goss 1993).

The morphology and the kinematics of the extended Northern Arm
are well-modeled by dispersion ring as it appears, for example, 
in the eighth frame
(t = 0.175).  Rotating the frame clockwise by 80 degrees
in the orbital plane, tilting the positive y-axis forward by
50 degrees (the inclination) and rotating the negative
x-axis (the line-of-nodes)
to a position angle of 35 degrees with respect to north, the structure
appears as in Fig.\ 8a where again arrows indicate the sense and 
magnitude of proper motion.  This projected dispersion ring,
at its northern most extent, appears rather thick in comparison
with the observed ionized filament.  This is consistent with the 
Northern Arm being the inner ionized edge 
of a broader contiguous region of neutral and molecular gas, the 
so-called ``Northern Intruder''
(Davidson et al. 1992, Jackson et al. 1993).

A contour map of the projected density is shown in Fig.\ 8b.  
Here it is evident that the eastern section of the
ring is more dense and, if ionized, would have a significantly higher 
emission measure than the more tenuous western side.  This is consistent
with the fact that the extended Northern Arm appears as
an incomplete elliptical ring;  i.e.,no western component 
is clearly observed.
Comparing Fig.\ 8b with Fig.\ 6 (the scale is the same) we see  
that the morphology of the Northern Arm is generally consistent with
dispersion ring model.  The points in Fig.\ 8b indicate the positions
where the radial velocity has been measured in Br$\gamma$ line emission
by Herbst et al. (1993);  again the coincidence with the projected 
dispersion ring is reasonable.

The radial velocity field is shown as a contour map in Fig.\ 8c.
This agrees with the run of radial velocity observed in
the Ne$^+$ line (Serabyn et al. 1988) as well as in infrared and
and radio hydrogen recombination lines 
(Herbst et al. 1993, Roberts et al. 1996) 
Another representation of the velocity field of this structure
is shown in Fig.\ 9 which is the radial velocity at the inner edge
as a function of position angle (as in Fig.\ 4 for the CND).  Also 
shown are the observations of the extended Northern Arm radial velocity
given by Herbst et al. (1993).  The evident agreement
lends credibility to this scenario.

It is perhaps more than fortuitous that the projection of this dispersion
ring giving the closest agreement with both kinematics and morphology
of the extended Northern Arm implies the orbital plane of this feature
almost coincides with that of the CND; the angle between the two planes
is 10 degrees.  
Here, in both cases, it is assumed that the northwestern
edge is the near side (observationally, this issue remains
unclear, see Mezger et al. 1996)
An alternative model is possible with
southwestern side as the near side (the proper motion vectors 
would be oppositely directed to those shown in Fig.\ 8a),
but then the plane of the Northern Arm would not coincide with that
of the CND (the specific orientation of the CND is favored 
in this picture by the the model fit to the
morphological and radial velocity asymmetries).  The probable near 
coincidence of the two planes would suggest that
there is a rather close relationship between the extended Northern Arm
and the CND-- that both originated from low angular momentum material in the
same plane and possibly from the same accretion event.

It is also noteworthy that this structure can only be reproduced if
a point mass is present; if the central object has a mass less than 
$10^6$ $M_\odot$ then the tidal stretching into a long filament does not
occur.  Although this adds very little to earlier arguments
on the necessity of a massive object at the Galactic Centre, it is
also of interest that the deviation from $1/r^2$ attraction in the
central two parsecs due to the contribution of the extended stellar
system leads to strong shocks during the formation of
the precessing dispersion ring.  Such shocks are necessary to obtain
the compression  required for star formation in the near vicinity of
the black hole (as discussed in the next section).
This consideration limits the mass of
the central object to be less than about $3\times 10^6$ $M_\odot$;  
otherwise the potential is too near Keplerian and there are no 
strong shocks.

The only conspicuous structure which is not accounted for in this
picture is the Eastern Arm.  In fact, this structure 
cannot be in
the orbital plane of the extended Northern Arm if its motion is also orbital.
Perhaps this feature traces a separate infall event or is part of the
ionized ragged inner boundary of the CND (Serabyn 1988, Morris \& 
Serabyn 1996).

\section{Star formation in strong shocks}

Gas transported inward toward the black hole on highly elongated and
intersecting orbits, as in Fig.\ 7 (at t=0.1 for example) 
will form extremely strong shocks at the point of intersection.
The post-shock density in streams colliding with relative velocities in 
excess of 100 km/s could exceed the Roche limit (eq.\ 1) for star formation 
at a tenth of a parsec from the hole (Phinney 1988).
This then provides a natural mechanism for formation of the young
stars seen in the central parsec of the Galaxy.

The triggering of star formation in regions of strong gas compression 
is easily modeled in the simulation described here because the 
divergence of the velocity field can be estimated.  At time $k$ and
at the
location of the $i^{th}$ particle this is 
$$ -(\nabla \cdot v)_{ik} = {{|\Delta {\bf v}_{ik}}|\over{\alpha_k\sigma}}
\eqno(7)$$ where $ {\Delta {\bf v}_{ik}}$ is given by eq.\ 4.   
Unsurprisingly, the region of largest gas compression is very near
the black hole where streams intersect. 
We may assume that a gas particle at a given location becomes a star
if the negative velocity divergence (compression) at that location 
exceeds some arbitrary threshold.  Once a particle has been converted
into a star the dissipation is turned off, and thereafter its motion is
determined only by the force of gravity.

The cloud infall illustrated in Fig.\ 7 has been repeated using this
algorithm for star formation with the compression threshold for star 
formation set at 2000 km/s/pc.  Most of the star formation occurs during
the first few passages of the cloud when the long filament intersects itself
at a large angle. 
Snapshots of the gas and star distributions at
a time of $3.0\times 10^5$ years (corresponding to more than one precession
period for the dispersion ring) are shown in Figs.\ 10a and 10b 
respectively.  By this epoch the gas has settled into the stable dispersion
ring and the star formation has almost ceased.  Roughly half of the original
4000 gas particles have been converted into stars.
The star and gas distribution at $5.5\times 10^5$ years,
almost one precession period later, are shown in Figs.\ 10c and 10d.
The gaseous dispersion ring, as seen in Fig.\ 10c, is clearly present 
after two complete precession periods;  it is a long lived and
stable feature which can persist for more than a million years. 
The morphology and kinematics of this feature continue to resemble
that of the extended Northern Arm for 10 to 20 orbit times.

Fig.\ 10 illustrates the effect of gas dissipation on the resulting 
morphology.  The fact that strong dissipational forces arise when
high velocity gas streams intersect gives rise to the  
coherent structure of the dispersion ring 
in which dissipation is minimized;  in the language of modern
dynamics, the dispersion ring appears as an ``attractor'' in the 
phase space of the system.   The ensemble of stars, with no
such mechanism for self-organization, slowly disperses throughout the
available phase space.
But, significantly, the time scale for phase mixing of the stellar orbits
is rather long.  This is because the stars are formed from the
gaseous dispersion ring--  i.e., in the 4-dimensional phase space
the domain of initial conditions 
with which these particles are created is quite restricted.
The rosette pattern of the stellar orbits is still evident as an 
underlying skeletal structure (Fig.\ 10b) after one complete precession 
period of the dispersion ring or about 5 orbit times at one parsec.  
It takes 2 or 3 precession periods for the stellar orbits to become 
well mixed (Fig.\ 10d).

This 
scenario, therefore, can provide an explanation not only for the existance
of young stars near a massive black hole but also for a spatial 
distribution of these stars which is not yet completely relaxed.  
Although stars can be formed a few tenths of a parsec from
the black hole, the relevant time scale is not that of an orbital
period in this small region, as previously supposed 
(Allen \& Sanders 1986, Sanders 1992),
but the precession period of the more extensive gaseous
dispersion ring from which these stars have been formed.  As is evident
from Fig.\ 10d the 
phase mixing time scale for the stellar orbits is in excess of 
two precession periods of
the dispersion ring-- in this case more than half a million years--
which is 100 times longer than the orbital period in the inner
few tenths of a parsec.  As the gas dispersion ring precesses it 
decays into a more axisymmetric torus on a time scale exceeding
a million years.  

Some of the young stars currently observed in the inner few 
tenths of a parsec may well have been formed quite recently out of
the tidal debris traced by the observed filaments.  However, given
that the likely lifetime of the youngest stars is in excess of 
one-million years, many of these stars may have also been formed
in recent previous infall events.  Gas clouds with low angular 
momentum do not necessarily  share the sense of Galactic rotation.
Therefore, the fact that the youngest stars appear to possess a
component of systematic counter-rotation (Genzel et al. 1996), 
may just reflect the 
direction of the original angular momentum vector of the tidally disrupted 
cloud from which these stars have formed.

\section{Conclusions}

It is now difficult to dispute
the probable presence at the Galactic Centre of a black hole with a mass
in excess of $10^6$ $M_\odot$.  However, as has been pointed out many
times before, the absence of the sort of activity normally associated 
with active galactic nuclei would imply that the Galactic Centre
black hole is currently in a quiescent phase-- that activity associated
with massive black holes in galactic nuclei is episodic, and, by implication,
that accretion in galactic nuclei is also episodic.

A schematic scenario of episodic accretion is that of a nucleus containing
a black hole and a few dozen massive clouds with random rather than 
systematic motion (Sanders 1981).  Occasionally a cloud directly
encounters the black hole leading to Bondi accretion and the buildup
of a short-lived accretion disk.  This picture was somewhat refined
by Bottema and Sanders (1986) who showed that in such an encounter, the
amount of material captured and the time scale for subsequent accretion
could fuel a short ($10^5$ year) outburst of Seyfert activity followed by  
10 million years of inactivity.

The observations of the circumnuclear material in the Galactic
Centre lend some support to this picture, although accretion events are
not so simple as a direct encounter of a molecular cloud with a black
hole.  It appears that accretion actually proceeds through the
tidal disruption and capture of clouds on low angular momentum orbits.
This can lead to the formation of structures resembling the CND and the
ionized extended Northern Arm in the central cavity of the CND.  
The general pattern
of such capture in the presence of a point mass
is the tidal stretching of a cloud into a long 
filament which collides with itself before being organized into a
long-lived, elliptical precessing dispersion ring.  If the specific angular 
momentum of the captured cloud is low enough the 
gas streams intersect with high velocity which almost certainly 
leads to very active
star formation even in the near tidal field of the black hole.
Thus the accretion process is highly inefficient in the sense that much
of the captured material forms stars rather than being accreted by
the black hole.

All of this requires a very clumpy and turbulent
interstellar medium in the inner few hundred parsecs of the Galaxy.
But this is not an assumption because molecular line observations
clearly reveal the presence of such a medium (see Morris \& Serabyn 1996 and
Mezger et al. 1996).  Of course one might reasonably ask what
provides a continuous supply of low angular momentum clouds and what
supports the highly supersonic turbulence.  Dynamical friction is one 
mechanism which can
move massive molecular clouds from 200 pc inward to the centre
on time scales less than $10^9$ years 
(Stark et al. 1991).  High turbulent velocities may be maintained by
the continuing star formation and by the occasional flare-up of the
black hole.  It is likely that the low specific angular momentum of 
the clouds forming the CND and ionized filaments results from
cloud-cloud collisions in the inner 5 to 10 pc.

In summary, the dispersion ring model
has many of the observed properties of the CND-- for example, an
asymmetric central cavity with the gas density being much higher on
the side nearest to the point mass at the dynamical centre.
In addition, at least some of the ionized gas filaments in the central parsec
(the extended Northern Arm) may also be identified with a highly elliptical
dispersion ring.  For a cloud with such low specific angular momentum,
on the way to becoming a
dispersion ring self-intersection of gas streams with high velocity
will probably lead to star formation, and this could well be the origin
of the very young massive stars actually observed in the central parsec
of the Galaxy.  Because the stars form from gas on a 
dispersion ring extending
from 0.1 to 1 pc and occupying a restricted domain of the available
phase space, phase mixing occurs on a time scale which is 
much longer than the
characteristic orbit time in the inner few tenths of a parsec
($10^6$ years instead of $10^4$ years).  Thus both the formation and 
unrelaxed distribution
of stars in the near tidal field of the black hole may be understood in
terms of this mechanism.

Many of these ideas are not new: i.e., the transient nature of the CND,
the identification of the ionized filaments with tidally stretched
clouds, star formation in strong shocks in colliding gas streams.
The new aspect suggested by the sticky particle code applied in this work
is the identification of the gaseous circumnuclear material with
asymmetric elliptical dispersion rings formed by
the tidal debris from clouds on low angular momentum orbits. 
The objective here is not to model precisely the morphology and kinematics of
the CND and the ionized filaments;  the parameter space of 
initial conditions is too large for that 
exercise to be particularly meaningful.  
Nonetheless, such models for the gas features seen in the inner few 
parsecs do make some predictions.  For example, consistency with
the morphological
and kinematic asymmetries of the CND (the dense inner boundary to the west,
the outer extention to the south, the ridge of positive
velocity to the west of \sgast)
gives preference to a specific spatial orientation:
the northwestern side is the near side.
Then, because the best fit dispersion ring model of the extended
Northern Arm suggests that the plane of this feature 
coincides with that of the CND, the proper motion
of individual gas clumps near \sgast, if this could be measured, would
have the directions shown in Fig.\ 8a; i.e., motion along the Northern
Arm is toward \sgast.  Moreover, we might expect that the
extended Northern Arm is a more complete elliptical structure
(Fig.\ 9) with a faint western counterpart.  
Higher sensitivity recombination line observations might unveil the
complete dispersion ring.  For now, the overall
similarity between the model and the observed morphology and 
kinematics of the CND and extended Northern Arm supports the 
plausibility of this scenario.

I am grateful to E. Hartlief for initial three-dimensional calculations
of tidal disruption which demonstrate the validity of the 
two dimensional calculation applied here.

\clearpage
\begin{figure}
\vspace{1.0cm}
\caption{ The circumnuclear disk as seen in HCN emission integrated
over radial velocity reproduced from the paper of G\"usten et al. (1987).
The heavy solid contours are in intervals of 0.12 K averaged over
300 km/s.  Also shown (light contours) is the 5 Ghz radio continuum
map by Lo \& Claussen (1983) illustrating the ionized filaments.  The
principal morphological features are labeled.  The arrows on the axes
indicate the position of the origin (\sgast)}
\label{Fig. 1}
\end{figure}

\begin{figure}
\vspace{1.0cm}
\caption {A sticky particle simulation of the tidal disruption of
a low angular momentum cloud near the Galactic Centre.  The clumpy
cloud is initially at a distance of 6.5 pc from the galactic centre
and moving tangentially with a velocity of 40\% that required for
a circular orbit.  The different frames (left to right) show the
passage of the cloud in the orbital plane.  The units of length
are parsecs with the origin being at the position of the  
$2.5\times 10^6$ $M_\odot$ point mass.  The elapsed time in 
millions of years is indicated for each frame.  The elliptical dispersion
ring with an asymmetric central cavity is evident after about half
a million years.  Over the next $5\times 10^5$ years the overall
shape is maintained but the central cavity gradually closes.  The CND
can be well-represented by the tidal debris as it appears 
in the seventh frame (t=0.85).}
\label{Fig. 2}
\end{figure}

\begin{figure}
\vspace{1.0cm}
\caption {  These three frames show the CND model at 
time t=0.85 projected onto the 
plane of the sky.  The offset from the point mass is given 
in seconds of arc assuming that the Galactic Centre is at a distance
of 8.5 kpc (north is up, east is to the left).  
The structure in the seventh frame of Fig.\ 2
has been rotated clockwise by 130 degrees, inclined with respect to the
plane of the sky by 60 degrees, and the line-of-nodes (shown by the
solid line) is rotated to 
a position angle of 35 degrees with respect to the vertical (north).
These are typical projection parameters for the CND where the 
northwestern side is the near side.  The scales are identical to
that of Fig.\ 1 and may be compared directly.  The frames show:  a) A particle
plot where the arrows at the
position of the particles indicate the sense and magnitude of the
velocity projected onto the plane of the sky (proper motion).  b) A contour
map of the gas surface density.  Units of density are arbitrary,
but the contour levels are in equal intervals of 12.5 \% of the maximum
density.  c)  A contour map of the radial velocity where solid contours
are positive radial velocity (away from the observer) and the dashed 
contours are negative values.  The contours are given in intervals of 
40 km/s.  Note the positive velocity ridge to the west of \sgast.}
\label{Fig. 3}
\end{figure}

\begin{figure}
\vspace{1.0cm}
\caption {The radial velocity at the inner edge of the projected
dispersion ring seen in Fig.\ 3 as a function of position angle
is shown here by the solid line.
Also shown by the unconnected points are the observations 
of HCN velocities at the inner edge of the CND by G\"usten et al. (1987).}
\label{Fig. 4}
\end{figure}

\begin{figure}
\vspace{1.0cm}
\caption{ The ionized filaments as seen in the integrated Ne$^+$ 
emission reproduced from Lacy et al. (1991).  This illustrates the 
ionized filaments in greater detail.  Again the principal features
are labeled and the dots indicate the extended Northern Arm}
\label{Fig. 5}
\end{figure}

\begin{figure}
\vspace{1.0cm}
\caption{ Contour map of the 12.4 $\mu$m continuum emission
of hot dust 
reproduced from Gezari et al. (1996).  The extended Northern Arm is the 
most conspicuous feature appearing as a partial ellipse wrapping around
\sgast.  The fainter Eastern arm appears as a morphologically 
distinct feature in this map.}
\label{Fig. 6}
\end{figure}

\begin{figure}
\vspace{1.0cm}
\caption {A time sequence of the evolution of tidal debris resulting
from the passage of a small cloud (r=0.4 pc) passing within 0.1 pc
from the centre of the gravitational field described by eq. 3.  
Initially the cloud centre is 2.4 pc from the point and 
moving tangentially with a velocity of 20\% that required for a 
circular orbit.  The distance with respect to \sgast is in units of 
parsecs and the time since
the beginning of the infall is given in units of one million years.
The cloud is stretched into a long filament which collides with
itself at high velocity several times before forming an
asymmetric, elliptical dispersion ring.  The structure precesses
in a sense counter to rotation with an angular frequency of 
26 km/s/pc and persists for many rotation periods.  The final frame
(t=0.185) is a reasonable representation of the extended Northern Arm.}
\label{Fig. 7}
\end{figure}

\begin{figure}
\vspace{1.0cm}
\caption {Three views of the dispersion ring model of the extended
Northern Arm at t=0.185, the final frame of Fig.\ 7,
projected onto the plane of the sky.  The structure as it appears
in Fig.\ 7 is rotated clockwise by 80 degrees,
inclined with respect to the sky by 50 degrees, and the line-of-nodes 
(solid line) is
rotated to a position angle of 35 degrees with respect to north.
As for the CND model, the northwestern side is 
the near side.  The offsets from the dynamical centre are given in seconds
of arc, and the large
X marks the position of the pont mass.  The scales are the same as in 
Fig.\ 6 and can be directly compared.
As in Fig.\ 3, the frames show:  a)  The direction and magnitude
of the gas motion projected onto the plane of the sky.   
b)  A contour map of the gas surface density. 
The units of density are
arbitrary but the levels are in equal intervals of 20\% of
the peak density.  The points
indicate the positions where the radial velocity has been measured
in Br$\gamma$ line emission by Herbst et al. (1993).  c) A 
radial velocity contour map.  The solid contours are levels
of positive velocity in intervals of
40 km/s  and dashed contours are of negative velocity also in 
intervals of 40 km/s.}
\label{Fig. 8}
\end{figure}

\begin{figure}
\vspace{1.0cm}
\caption{The radial velocity of the dispersion ring model
for the extended Northern Arm 
as a function of position angle (with respect to \sgast)
is shown by the solid line.  
Also shown by the unconnected points are the observations
of Serabyn \& Lacy (1985) and Herbst et al.\ (1993).}
\label{Fig. 9}
\end{figure}

\begin{figure}
\vspace{1.0cm}
\caption {The distributions of gas and stars at two different
epochs of the Northern Arm simulation.  The distance offset from \sgast
is given in parsecs.  In this simulation stars
are allowed to form in regions of strong compression;  when a gas
particle is converted into a star, the viscous dissipation is turned
off and the motion is subsequently determined only by the gravitational
force.  Frames $a$ and $b$ show the gas and stars at t=0.30 corresponding
to more than one complete precession of the dispersion ring.  The
gas exhibits the clear structure of the dispersion ring.  The stars
are phase mixing but are still asymmetrically distributed and reflect
the rosette orbit structure.  Frames $c$ and $d$ show the gas and stars
at t=0.55, almost one precessional period later.  The gas still exhibits the
organized structure of the dispersion ring but the stars are
now distributed more symmetrically with respect to the centre.}
\label{Fig. 10}
\end{figure}

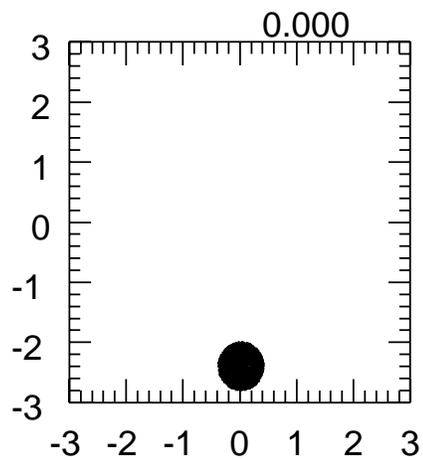
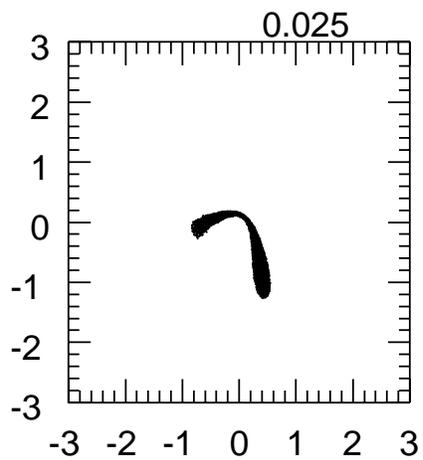
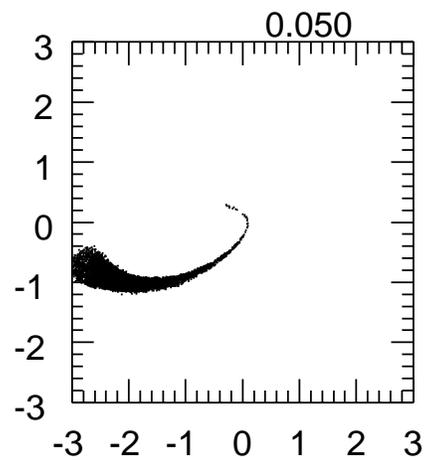
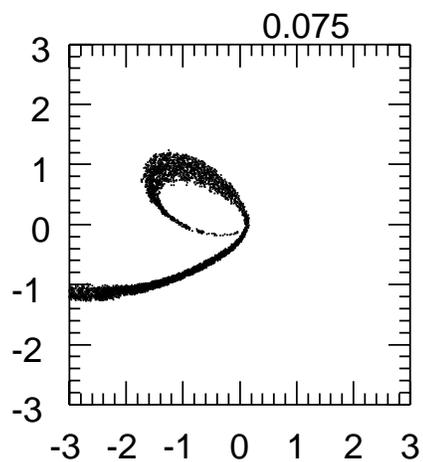
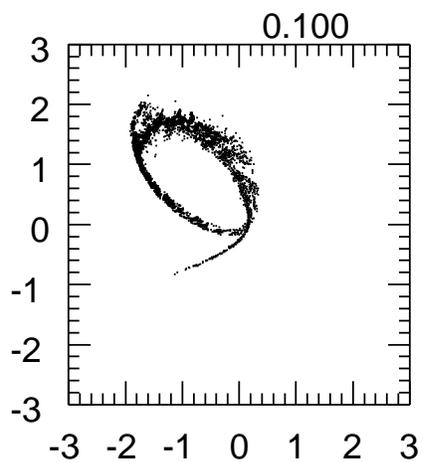
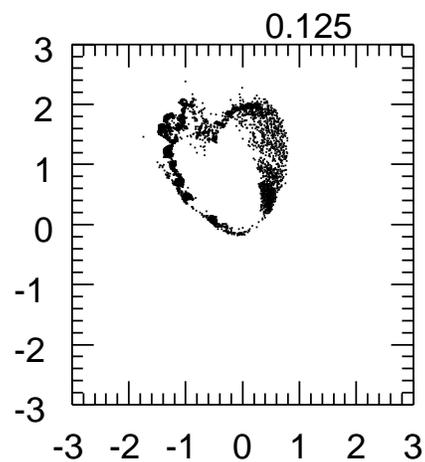
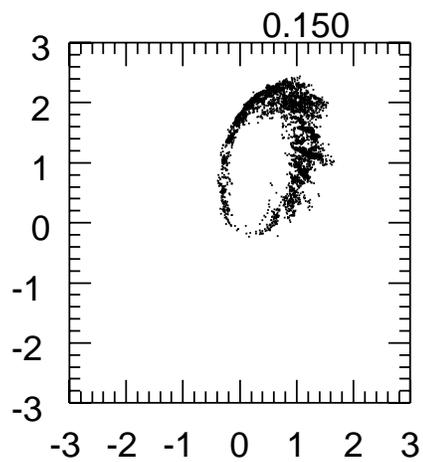
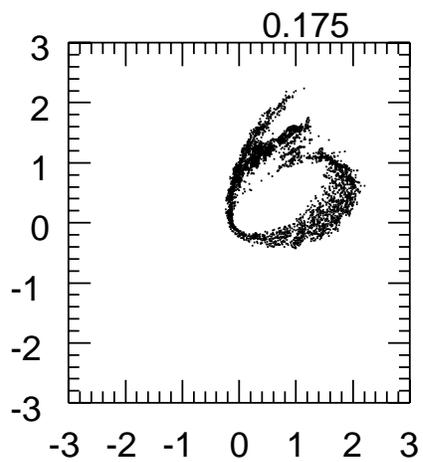
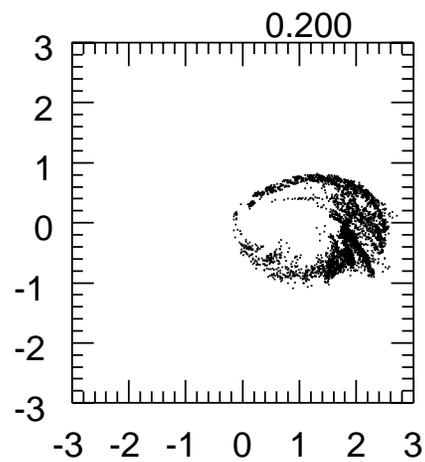
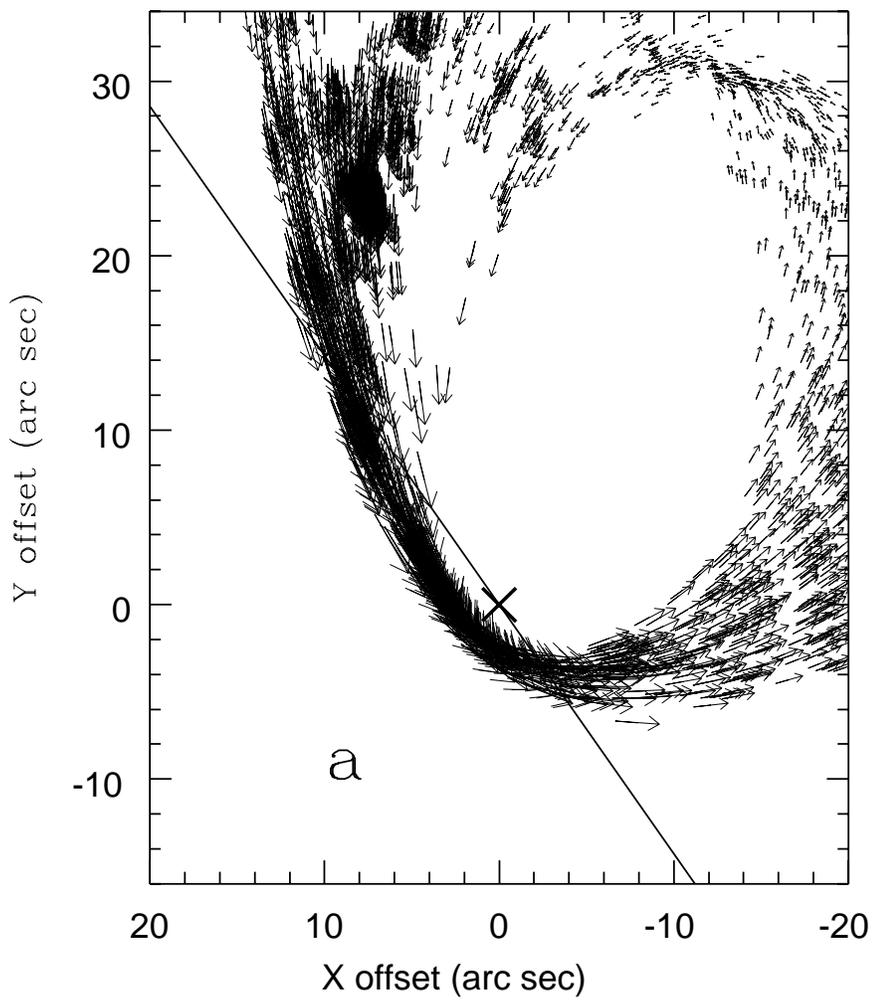
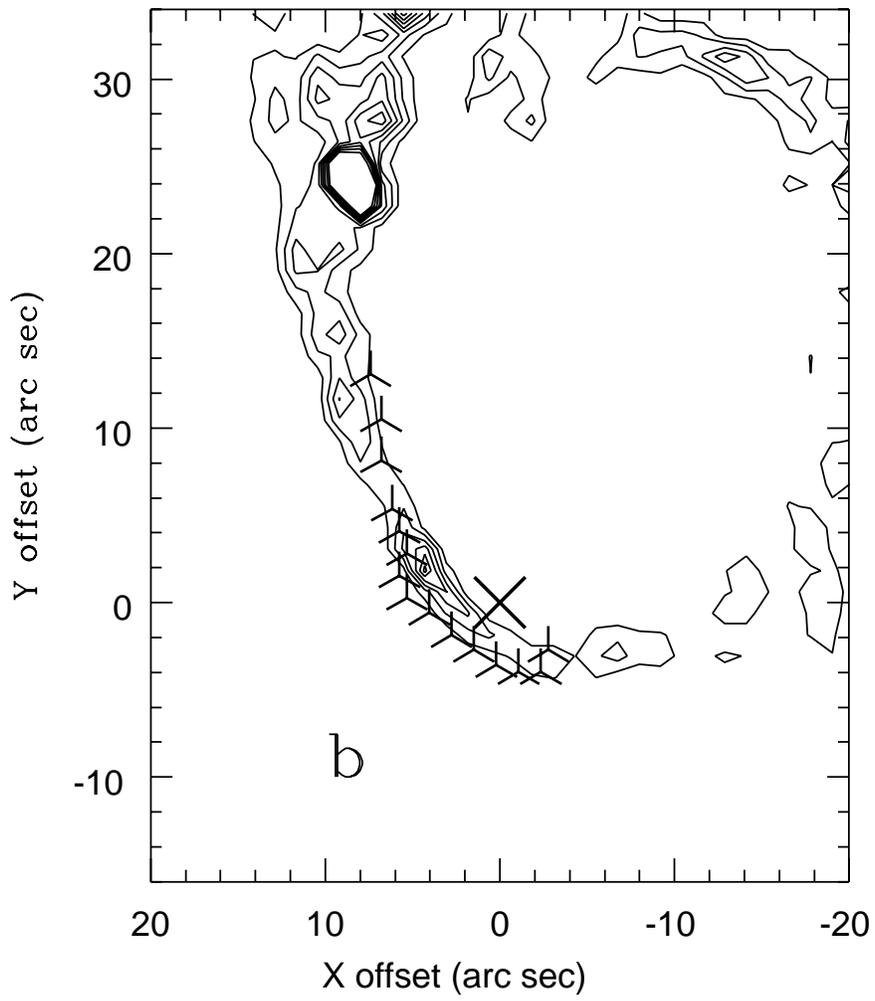
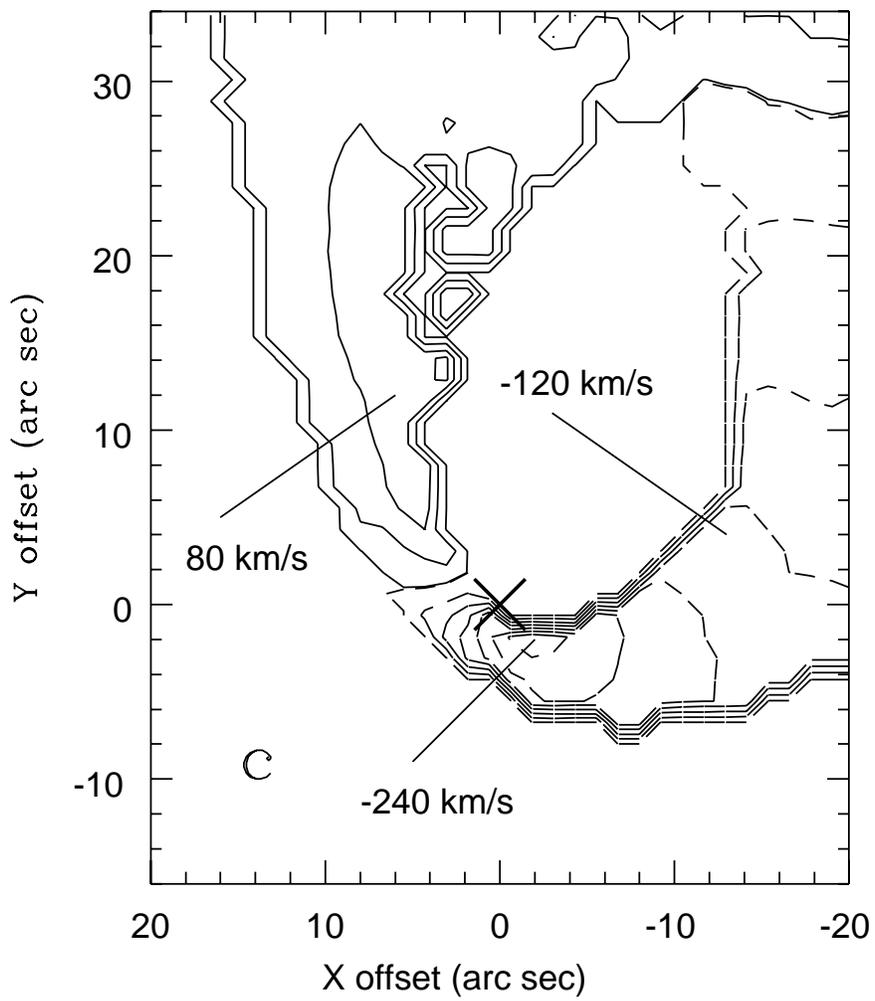
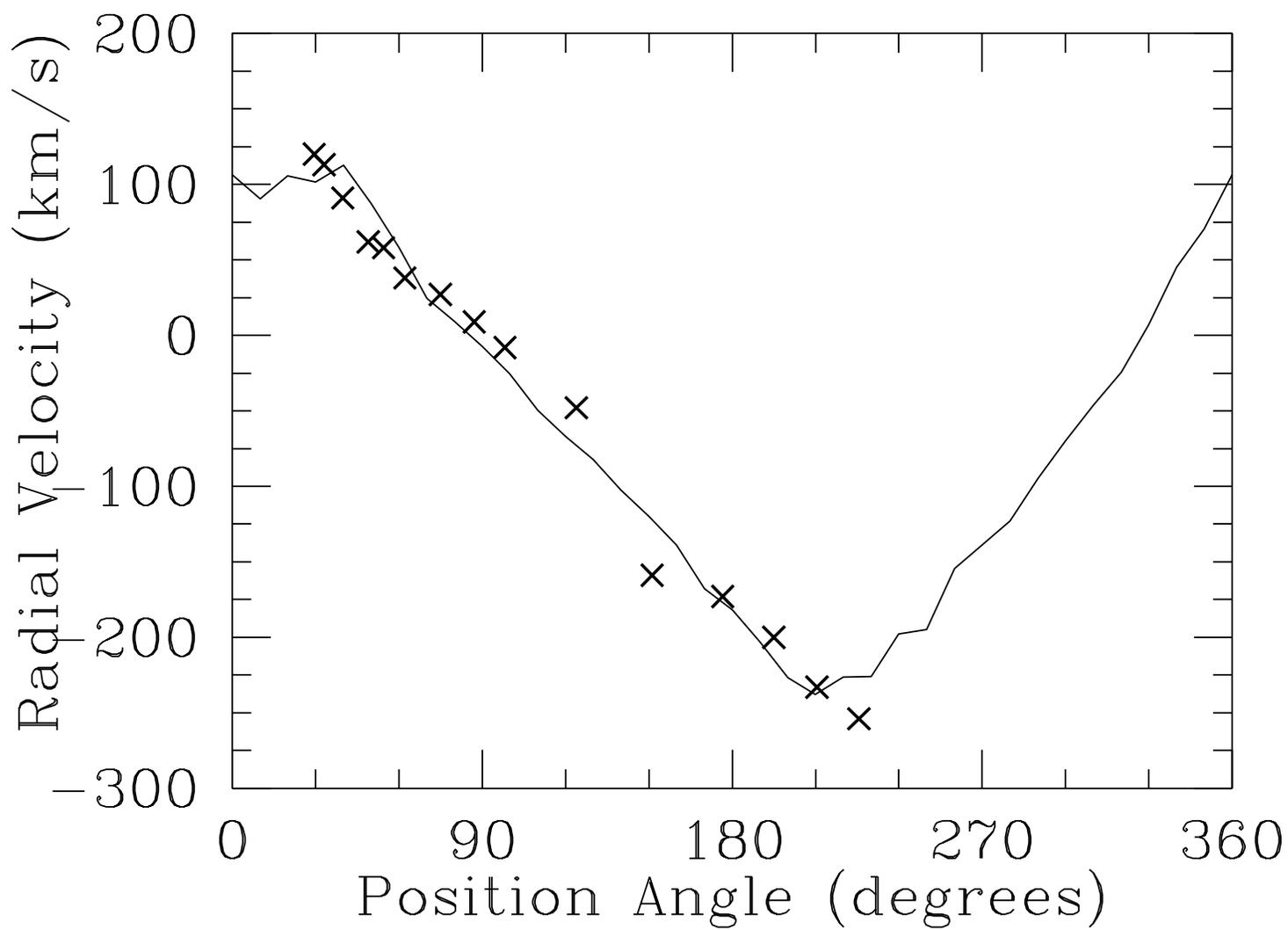
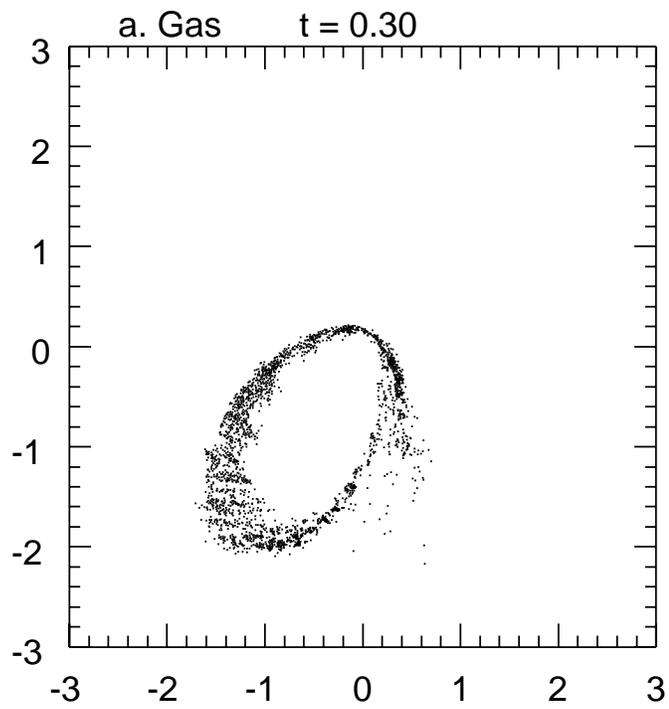
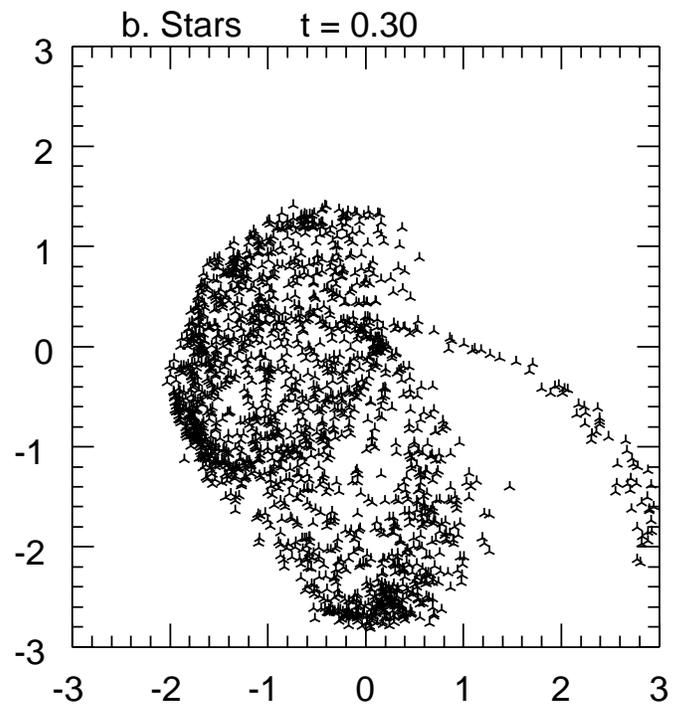
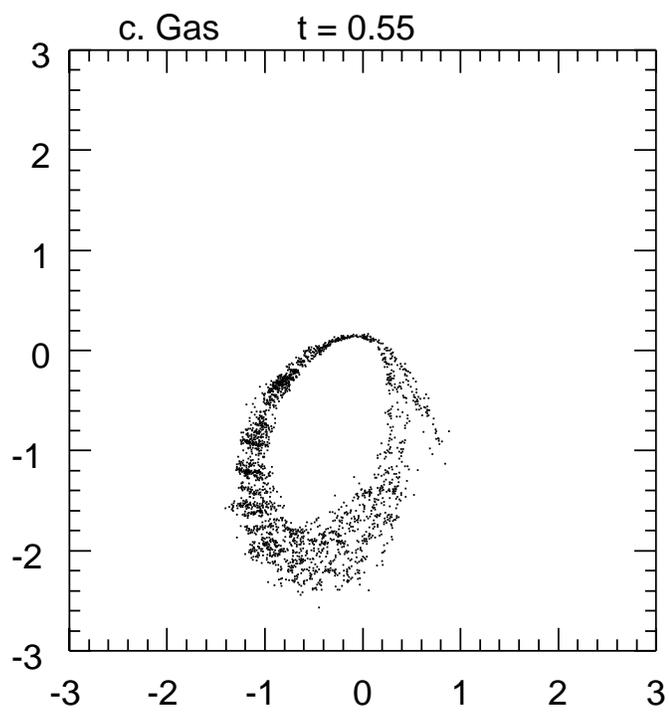
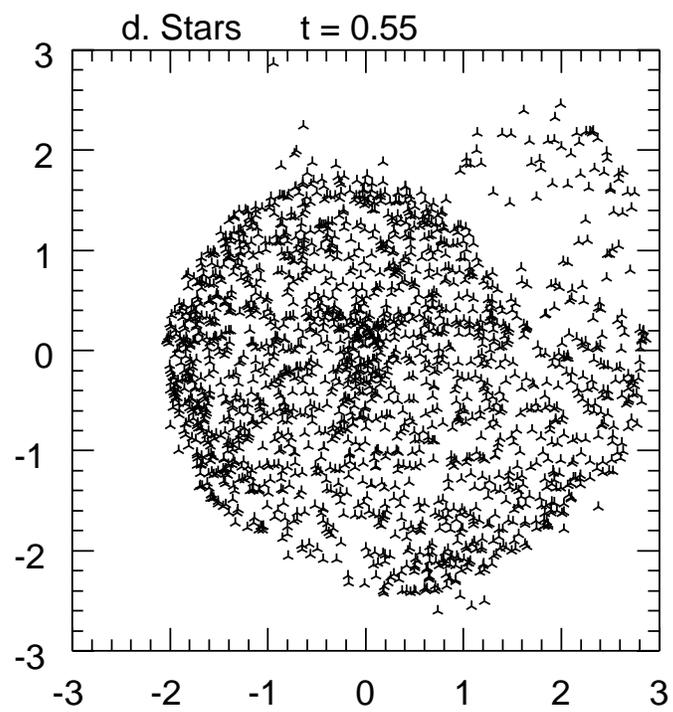
\end{document}